\documentstyle[aps,epsfig,prl]{revtex}
\begin{document}
\title{Transmission of a Symmetric Light Pulse through a Wide QW}
\author{ L. I. Korovin, I. G. Lang}
\address{A. F. Ioffe Physico-technical Institute, Russian
Academy of Sciences, 194021 St. Petersburg, Russia}
\author{D. A. Contreras-Solorio\dag, S. T. Pavlov\dag\ddag}
\address{\dag Facultad de Fisica de la UAZ, Apartado Postal C-580,
98060 Zacatecas, Zac., Mexico \\
\ddag P. N. Lebedev Physical Institute, Russian Academy of Sciences, 119991 Moscow,
Russia}

 \twocolumn[\hsize\textwidth\columnwidth\hsize\csname @twocolumnfalse\endcsname
\date{\today}
\maketitle\widetext
\begin{abstract}
\begin{center}
\parbox {6in}
{The reflection, transmission and absorption of a symmetric electromagnetic pulse,
which   carrying frequency is close to the frequency of an interband transition in a
QW (QW), are obtained. The energy levels of a QW are assumed discrete, one exited
level is taken into account. The case of a wide QW is considered when a length of the
pulse wave, appropriate to the carrying frequency, is comparable to the QW's width
and it is necessary to take into account a dependence of a matrix element of an
interband transition from the wave vector of light. Alongside with it the distinction
in parameters of refraction of substances of a QW and barrier is taken into account.
The formulas for electrical fields on the right and to the left of the QW,
appropriate reflected and transmitted through a well pulses at any ratio between
radiative and non-radiative broadenings of the exited energy level of the electronic
system, are obtained. In figures the time dependencies of the dimensionless
reflection, absorption are transmission (which are defined as modules of the
appropriate energy fluxes  and their sum is equal to the module of a flux of
 an exciting pulse energy) are represented. It is shown, that the spatial
dispersion and a distinction in refraction indexes influence stronger reflection, as
alongside with reflection connected with interband  transitions in the QW, the
additional reflection from borders of the QW takes place. In comparison with model,
in which the matter  is considered homogeneous and the spatial dispersion of light is
not taken into account,the most radical changes take place in reflection in a case,
when the radiative broadening  of the exited state is large in comparison with non-
radiative broadening. On the side of the large values of the QW's width the theory is
limited by a condition of existence of  size-quantized energy levels.}
\end{center}
\end{abstract}

] \narrowtext

\section{Introduction}
 Lately in \cite{1,2,3,4,5,6,7} the change of the light pulse form was investigated
at its transmission through a QW.  Asymmetric exciting pulses with abrupt fronts
\cite{1,2,3} and symmetric pulses \cite{4,5} have been considered. It was supposed,
that the exciting pulse carrying frequency   $ \omega _ {\, l \,} $ is close to the
electronic excitation frequency  $ \omega _ {\, 0 \,} $ (two-level system)
\cite{1,2,5}.  A three-level system \cite{7} and system with many exited states
\cite{3,6} are investigated also. Results of these works are true for rather narrow
QW's, when the inequality
\begin{equation}
\label{1} \kappa \, d\ll 1
\end{equation}
is carried out, where $d $ is the QW's width, $ \kappa $ is the module of a light
wave vector appropriate to the carrying frequency of the symmetric light pulse.
Actually the parameter $ \kappa \, d $ in mentioned works relied equal to zero and
calculated there reflection, absorption and transmission did not depend on the QW's
width. For a numerical estimation of size $ \kappa $ we use  a wave length of
radiation of a hetero-laser on a basis GaAs, equal $ 0.8 \mu $. The energy,
corresponding  to this  wave length, is as follows $ \hbar \, \omega _ {\, l \,} =
1,6 \, eV $. If a refraction index of a QW's matter is $ \nu =3,5 $, $ \kappa = \nu
\, \omega _ {\, l \,}/ \, c \, = 2,8.10^5 cm ^ {-1} $, where $c $ is the light
velocity in vacuum. For the QW width $d=500A $, the parameter $ \kappa \, d=1,4 $.
Thus, for wide QWs the spatial dispersion of waves, consisting an exciting pulse, can
appear essential.

For wide QWs an inequality $d\gg a _ {\, 0 \,} $, where $a _ {\, 0 \,} $ is the
lattice constant, is very strong and at a description of a pulse transmission through
a QW it is possible to use  Maxwell's equations for continuous matter. At such
approach it is necessary to take into account a distinction in refraction indexes of
barriers and QW. Then additional reflections should appear  from borders of the QW,
which decreases with  reduction of the parameter $ \kappa \, d $, but in an area $
\kappa \, d \geq 1 $ can in some cases  be compared or exceed reflection caused by
resonant transitions in a QW. Transmission of a light wave will change together with
reflection. Thus, alongside with the account of a dependence of reflection and
transmission on parameter $ \kappa \, d $ one should take into account distinction in
refraction indexes of barriers and QW. At present work the influence of these two
factors on the form of reflected and transmitted pulses is taken into account.

The system consisting from a deep semiconductor QW of the I type located in an
interval $0\leq z \leq d $ and two semi-infinite  barriers is considered. It is
supposed, that the exciting light pulse moves along an axis $z $ from negative $z $.
It is supposed also, that barriers are transparent for the pulse, and in the QW
 the pulse is absorbed, causing resonant interband transitions. An
intrinsic semiconductor and zero temperatures are meant. Exited states are taken into
account only, in which one electron  from the valence band transited in the
conductivity band, thus a hole appears in the valence band.  It is supposed, that $
\omega _ {\, l \,}\cong \omega _ {\, g \,} $ (the energy gap in the QW is $E _ {\, g
\,} =\hbar \, \omega _ {\, g \,} $) and a small part of valent electrons, located
close to the valence band extremum (for which the method of effective mass is true)
participates in absorption. For deep QWs in this case it is possible to neglect
tunneling of electrons in a barrier and to consider, that electrons are absent in
 barriers. Besides it is possible to consider energy levels located close to the
QW's bottom in approximation of indefinitely deep QWs. Considered system is
inhomogeneous. As for wide QWs the inequality (1) is not performed, optical
characteristics of such system are necessary to determine from the solution of
Maxwell's equations, in which current and charge densities,
 following from a microscopic consideration, are used \cite{8,9}.

Final results will be obtained for only discrete electron energy level
 in a QW. Influence of other energy levels on light reflection and
absorption  may be neglected, if the carrying frequency $ \omega _ {\, l \,} $ is
close to the excitation frequency of the chosen level $ \omega _ {\, 0 \,} $, and
other energy levels are located far away from the chosen energy level. Discrete energy
levels in a QW in a case $ \hbar \, {\bf K} _ {\, \perp \,} =0 $ (where $ {\bf K} _
{\, \perp \,} $ is the  total wave vector of the
 electron-hole pair (EHP)  in a QW's plane) are excitonic
energy levels in zero magnetic field, or energy levels in quantizing  magnetic field
directed perpendicularly to the QW's plane. As a convenient example of an energy level
we consider below the EHP energy level in a quantizing  magnetic field directed along
axis $z $, without taking into account the Coulomb interaction between electron and
hole, which is a weak perturbation  for strong magnetic fields and not wide QWs
\cite{10}. However, the excitonic effect will not result into basic changes  of our
results, but will affect only the magnitude  of radiative broadening $ \gamma _ {\, r
\,} $ of the electronic excitation in a QW. The same concerns and to excitonic levels
in a zero magnetic field.

\section{The Fourier transform of the electric field induced by an exciting light pulse.}

 The symmetric exciting pulse, to which
corresponds  electrical field of circular polarization, falls on a single QW from
negative $z $
\begin{eqnarray}
\label{2} { \bf E}_{\, 0 \,} (z, t) = {\bf e}_{\,l\,}E_{\,0\,}e^{-i\,\omega_{\,l\,}p}
\{\Theta (p) e ^ {-\gamma_{\, l \,}
p/2}\nonumber\\
 + [1-\Theta (p)] e^{\gamma _ {\, l \,} p/2} \} + c. c..
\end{eqnarray}
Here $E _ {\, 0 \,} $ is the real amplitude, $p=t-\nu _ {\, 1 \,} \, z/ \, c $,
\begin{equation}
\label{3} { \bf e}_{\, l \,} = ({\bf e}_{\, x \,}\pm i{\bf e}_{\, y \,})/{\sqrt 2}
\end{equation}
is the unit vector of circular polarization, $ {\bf e} _ {\, x \,} $ and $ {\bf e} _
{\, y \,} $  are unite vectors, $ \nu _ {\, 1 \,} $  is the barrier refractions
index, $c $ is the light velocity in vacuum, $ \Theta (p) $  is the Haeviside
function, $ \gamma _ {\, l \,} $ determines the  increasing and attenuation of the
symmetric pulse. The Fourier transform  of the function $ {\bf E} _ {\, 0 \,} (z, t)
$ looks like
\begin{eqnarray}
\label{4}  {\bf E} _ {\, 0 \,} (z, \omega) = \exp (i\kappa _ {\, 1 \,} z) \{{\bf e} _
{\, l \,} E _ {\, 0 \,} (\omega) + {\bf e} _ {\, l \,} ^*E _ {\, 0 \,} (-\omega) \},
\nonumber\\
 E _ {\, 0 \,} (\omega) =E _ {\, 0 \,}\gamma _ {\, l \,} / [(\omega -\omega
_ {\, l \,}) ^ 2 + ( \gamma _ {\, l \,}/2) ^2],
\end{eqnarray}
where $ \kappa_{\, 1 \,} =\nu _{\, 1 \,}\omega /c $.

In \cite{11} the task about transmission of a monochromatic electromagnetic wave
through a QW is solved with taking into account its spatial dispersion. The expression
for the density of a high-frequency current, induced by the transmitting
electromagnetic wave, was also obtained there. For the case of the only exited energy
level and circular polarization of falling waves the  current density looks like
\begin{eqnarray}
\label{5} \bar {\bf J} (z, t) = (1/2\pi) \int _ {-\infty} ^ \infty \, d \,\omega \exp
(-i \,\omega \, t) \bar {\bf J} (z, \, \omega),\nonumber\\
\bar {\bf J} (z, \, \omega) = -\frac {{\bf e} _ {\, l \,}
\gamma _ {\, r \,} \,\nu \, \omega} {4\pi} \Phi (z)\nonumber\\
\times \left [\frac {1} {\omega-\omega _ {\, 0 \,} + i \,\gamma /2} + \frac {1}
{\omega +\omega _ {\, 0 \,} + i \,\gamma /2} \right] \nonumber\\
\times \int_0^ddz ^ {\, \prime} A (z ^ {\, \prime}, \omega) \, \Phi (z ^ {\, \prime})
+c.c. \, = \, {\bf e} _ {\, l \,} \,\bar {J} (z \, \, t),
\end{eqnarray}
where
\begin{eqnarray}
\label{6} \hbar \, \omega _ {\, 0 \,} =\hbar \, \omega _ {\, g \,} +\varepsilon \, (m
_ {\, v \,}) + \varepsilon \, (m _ {\, c \,})\nonumber\\
 +\hbar \, \Omega _ {\,\mu \,}
(n+1/2)
\end{eqnarray}
is the energy of the interband transition appropriate to the chosen exited state, $
\varepsilon \,(m_{\,c\,})\,(\varepsilon\,(m_{\,v\,}))$  is the size-quantized
electron (hole)energy  with the quantum number $ m _ {\, c \,} \, (m _ {\, v \,}) $, $
\Omega _ {\,\mu \,} = | e|H/\mu \, c $  is the cyclotron frequency, $e $  is the
electron charge, $H $  is the magnetic field intensity,
$\mu=m_{\,e\,}\,m_{\,h\,}/(m_{\,e\,}+m_{\,h\,})$, $m _ {\, e \,} \, (m _ {\, h \,}) $
 is the electron (hole) effective mass, $n $  is the Landau quantum number, $
\gamma $  is radiative broadening of the exited state, $ \nu $ - is the QW refraction
index. In approximation of an indefinitely deep QW
\begin{equation}
\label{7} \Phi (z) = (2/d) \sin (\pi \, m _ {\, c \,} z/d) \, \sin (\pi \, m _ {\, v
\,} z/d).
\end{equation}
In Eq. (5) radiative broadening $ \gamma _ {\, r \,} $ is introduced for the EHP in a
magnetic field at $ \kappa \, d=0 $
\begin{equation}
\label{8} \gamma _ {\, r \,} = (2e^2/\hbar \, c \,\nu) (p _ {\, c \, v \,} ^ 2/m _
{\, 0 \,} \hbar \, \omega _ {\, g \,}) ( |e|H/m _ {\, 0 \,} c),
\end{equation}
where $m _ {\, 0 \,} $  is the free electron mass. The scalar $ A (z, \omega) $,
connected with vector potential in the Fourier representation $ {\bf A} (z, \omega) $
is as follows
\begin{equation}
\label{9} { \bf A} (z, \omega) = {\bf e} _ {\, l \,} A (z, \omega) + {\bf e} _ {\, l
\,} ^ {*} A (z, -\omega).
\end{equation}
The formula, similar Eq. (9), takes place and for an electric field vector $ {\bf E}
(z, \omega) $. Eq. (5) is true for heavy holes in crystals with the zinc blend
structure , if axis $z $ is directed along of  4-th order symmetry axis \cite{12,13}.
Included in $ \gamma _ {\, r \,} $ the real constant $p _ {\, c \, v \,} $ is
connected with an interband matrix element of a momentum for two degenerated bands:
$$ {\bf p} _ {\, c \, v \,} ^ {I, II} =p _ {\, c \, v \,} ({\bf e} _ {\, x \,}
\mp i \, {\bf e} _ {\, y \,}) / {\sqrt 2}. $$ The current density  $ \bar {\bf J} _
{\, l \,} (z, t) $ satisfies to a condition $div \, \bar {\bf J} _ {\, l \,} (z, t)
=0 $ and, hence, the induced charge  density $ \rho (z, t) =0 $. Then it is possible
to use calibration $ \varphi (z, t) =0 $, where $ \varphi (z, t) $ is the scalar
potential, and
\begin{eqnarray}
\label{10} { \bf E} (z, t) = (-1/c) (\partial {\bf A} /\partial t),\nonumber\\ { \bf
E} (z, \omega) = (i \,\omega /c) {\bf A} (z, \omega).
\end{eqnarray}
Since  $ {\bf E} (z, \omega) \sim {\bf A} (z, \omega) $, instead of an equation for $A
(z, \omega) $ it is more convenient to solve the similar equation for the scalar $ E
(z, \omega) $, which looks like
\begin{eqnarray}
\label{11}
 d^{\,2\,}E(z,\omega)/dz^{\,2}\,+\kappa^2E(z,\omega)=-(4\pi/c)\bar J (z, \omega),
\nonumber\\ \kappa =\nu \, \omega/c,
\end{eqnarray}
where in expression for $ \bar J (z, \omega) $ Eq. (5) it is necessary to replace,
using (10), $ A (z ^ {\, \prime}, \omega) $ by $E (z ^ {\, \prime}, \omega)$. Eq. (11)
is integro-differential one. If to represent formally its solution as the sum of the
general solution of the homogeneous equation and partial solution of the
inhomogeneous equation, one obtains  instead of Eq. (11) the Fredgolm integral
equation of the second kind \footnote {\normalsize {The similar equation was examined
in \cite{14} for an inversive layer; in \cite{11} the exact solution of the Eq. (12)
for a case of the monochromatic exciting wave is obtained.}}
\begin{eqnarray}
\label{12}
 E (z, \omega) =C _ {\, 1 \,} e ^ {i \,\kappa \, z} +C _ {\, 2 \,} e ^ {-i \,\kappa \, z}
 \nonumber\\ -
 \frac{i\,(\gamma_{\,r\,}/2)F(z)}{\omega-\omega_{\,0\,}+i\,\gamma/2}
 \int_0^d \, d \, z ^ {\,\prime \,} \, E (z ^ {\, \prime \,}, \omega) \,
\Phi (z ^ {\, \prime \,}),
\end{eqnarray}
which is true for $ \omega $, close to $ \omega _ {\, 0 \,} $, as its conclusion in
Eq. (5) for $ \bar {\bf J} (z, \omega) $ it was not taken into account not resonant
composed $ \omega +\omega _ {\, 0 \,} + i \,\gamma/2 $.
 The neglect by not resonant term is equivalent to an inequality
 $(\omega-\omega_{\,0\,})/\omega_{\,0\,}\ll 1 $. Thus theory
 becomes inexact at $ \omega - \omega _ {\, 0 \,}\approx \omega _ {\, 0 \,} $,
 however, this area
of frequencies is located far away from the resonant frequency $ \omega _ {\, 0 \,} $
and
  does not represent any interest. In time representation discrepancy of the theory
 appears  on times $t\leq t _ {\, 0 \,} =\omega _ {\, 0 \,} ^ {-1} $.
 If $ \hbar \omega _ {\, 0 \,} = 1.6 eV $, $t _ {\, 0 \,} = 4.10 ^ {-16} sec $.
The arbitrary constants $C _ {\, 1 \,} $ and $C _ {\, 2 \,} $ are determined from
boundary conditions in planes $z=0 $ and $z=d $, and the function $F (z) $ looks like
\begin{eqnarray}
\label{13} F (z) =e ^ {i \,\kappa \, z} \int_0^z \, dz ^ {\,\prime \,} e ^ {-i
\,\kappa \, z ^ {\,\prime \,}} \, \Phi (z ^ {\,\prime \,}) \,\nonumber\\
 + \, e ^ {-i
\,\kappa \, z} \int_z^d \, dz ^ {\,\prime \,} e ^ {i \,\kappa \, z ^ {\,\prime \,}}
\, \Phi (z ^ {\,\prime \,}).
\end{eqnarray}
If $ \gamma _ {\, r \,}\ll \gamma $,  the integral term in Eq. (12) is possible to
consider as a small perturbation and then it is enough to take into account the first
approximation on the integral term. Radiative broadening of energy levels in
quasi-two-dimensional systems appears due to an infringement of translational symmetry
in a direction, perpendicular to the QW plane \cite{15,16}. In the case of high
quality QWs scattering on heterogeneities of QW's borders can give a small
contribution in non-radiative broadening of an energy level. The same concerns and to
scattering on phonons and impurities at low temperatures and small impurity
concentrations. In result it can occur, that $ \gamma _ {\, r \,}\geq \gamma $. In
this case in the solution of Eq. (12) it is impossible to be limited by the first
iteration, and it is necessary to summarize all the iterative sequence. It is
possible to show \cite{11}, that this sequence is reduced to a geometrical
progression and the solution can be written as follows
\begin{eqnarray}
\label{14} E (z, \omega) =C _ {\, 1 \,} e ^ {i \,\kappa \, z} +C _ {\, 2 \,} e ^ {-i
\,\kappa \, z}\nonumber\\ -
 \frac{i\,(\gamma_{\,r\,}/2)\,F(z)}{\omega-\omega_{\,0\,}+
 i \, (\gamma +\gamma _ {\, r \,}\varepsilon) /2}\nonumber\\
 \times \int_0^d \, d \, z ^ {\,\prime \,}
 (C _ {\, 1 \,} e ^ {i \,\kappa \, z ^ {\, \prime \,}} +
 C _ {\, 2 \,} e ^ {-i \,\kappa \, z ^ {\, \prime \,}}) \, \Phi (z ^ {\, \prime \,}).
\end{eqnarray}
The complex value $ \varepsilon $
\begin{equation}
\label{15}
 \varepsilon=\varepsilon^{\,\prime}+i\varepsilon^{\,\prime \, \prime} =
\int_0^d \, d \, z ^ {\,\prime \,} \Phi (z ^ {\,\prime \,}) F (z
^ {\,\prime \,})
\end{equation}
determines the broadening change and energy level shift, which occur due to the wave
spatial dispersion. In a limiting case $ \kappa \, d=0\quad \varepsilon = \delta _
{\, m _ {\, c \,} \, m _ {\, v \,}} $. In barriers, where the induced current is
absent, instead of Eq. (11) it is true the equation
\begin{eqnarray}
\label{16} d ^ {\, 2 \,} E (z, \omega) /d \, z ^ {\, 2 \,}+
 \,\kappa_{\,1\,}^{\,2\,}E(z,\omega)=0,\nonumber\\
 ( z\leq 0, \, z\geq d), \quad \kappa _ {\, 1 \,} =\nu _ {\, 1 \,}
\, \omega/c,
\end{eqnarray}
the solution of which looks like
\begin{eqnarray}
\label{17}
 E^{\,l\,}(z,\omega)=E_{\,0\,}(\omega)e^{i\,\kappa_{\,1\,} z} +
C _ {\, R \,} e ^ {-i \,\kappa _ {\, 1 \,} z}, \; (z\leq 0), \nonumber\\
 E^{\,r\,}(z,\omega)=C_{\,T\,}e^{i\,\kappa_{\,1\,}z},\;(z\geq d).
\end{eqnarray}
The first term in the expression for $E ^ {\, l \,} (z, \omega) $ is the scalar
amplitude of the Fourier-transform of the exciting pulse, $C _ {\, R \,} $ determines
the reflected  wave amplitude, $C _ {\, T \,} $ determines the the transmitted wave
amplitude past through a QW. The factors $C _ {\, 1 \,} $, $C _ {\, 2 \,} $, $C _ {\,
R \,} $ and $C _ {\, T \,} $,  being functions of the frequency $ \omega $, are
determined from the continuity conditions of $E (z, \omega) $ and $dE (z, \omega) /dz
$ on borders $z=0 $ and $z=d $. It results in
\begin{eqnarray}
\label{18} C_{\,1\,}=(2E_{\,0\,}(\omega)/\Delta)e^{-i\,\kappa \, d} [ 1 +\zeta +
(1-\zeta) {\cal N}], \nonumber\\
 C _ {\, 2 \,} =- (2E _ {\, 0 \,} (\omega) /\Delta)
(1-\zeta) [e ^ {i \,\kappa \, d} + {\cal N}],\nonumber\\
 C_{\,R\,}=E_{\,0\,}(\omega)\rho/\Delta),\nonumber\\
  C _ {\, T \,} =
 4E_{\,0\,}(\omega)\,\zeta\,e^{-i\,\kappa_{\,1\,}d}
[ 1+e ^ {-i \,\kappa \, d} {\cal N}] /\Delta;
\end{eqnarray}
\begin{eqnarray}
\label{19} \Delta = (\zeta+1) ^2e ^ {-i \,\kappa \, d} - (\zeta-1) ^2e ^ {i \,\kappa
\, d}\nonumber\\
 - 2 (\zeta-1) {\cal N} [(\zeta+1) e ^ {-i\kappa d} + \zeta-1],\nonumber\\
 \rho=2i (\zeta^2-1) \sin {\kappa \, d}\nonumber\\
  + 2 [(\zeta^2+1) e ^ {-i \,\kappa
\, d} + \zeta^2-1] {\cal N}.
\end{eqnarray}
In Eqs. (18), (19) the designations are entered
\begin{equation}
\label{20}
 \zeta=\kappa/\kappa_{\,1\,}=\nu/\nu_{\,1\,},
\end{equation}
\begin{equation}
\label{21} { \cal N} = -i \, (\gamma _ {\, r \,}/2) F ^ {\, 2 \,} (0) / [
\omega-\omega_{\,0\,}+i\,(\gamma+\gamma_{\,r\,}\varepsilon)/2].
\end{equation}
The function $E _ {\, 0 \,} (\omega) $ is determined by Eq. (4). It follows from Eqs.
(13), (15), that in the case $m _ {\, c \,} = m _ {\, v \,} = m $ (an admitted
interband transition in a limit $ \kappa \, d=0 $)  $F (z) $ and $ \varepsilon $ are
equal:
\begin{eqnarray}
\label{22} F (z) =iB [2-\exp (i \,\kappa \, z) -\exp (i \,\kappa \, (d-z))\nonumber\\
- ( \kappa \, d/\pi m) ^2\sin^2 (\pi \, m \, z/d)],
\end{eqnarray}
\begin{eqnarray}
\label{23} F (0) =F (d) =iB [1-\exp (i \,\kappa \, d))], \nonumber\\
 B =
(4\pi^2m^2/\kappa \, d) / [4\pi^2m^2- (\kappa \, d) ^2],
\end{eqnarray}
\begin{eqnarray}
\label{24} \varepsilon^{\,\prime\,}=F^2(0)\exp(-i\,\kappa \, d) =
4B^2\sin^2 (\kappa \, d/2),\nonumber\\
 \varepsilon^{\,\prime\,\prime\,}=2B[1-B\sin{\kappa \, d} -
3 (\kappa \, d) ^2/8\pi^2m^2].
\end{eqnarray}
In the Fourier-representation the electrical field vector $ {\bf E} ^ {\, r \,} (z,
\omega) $ to the right of a QW in agreement with Eq. (17) looks like
\begin{eqnarray}
\label{25} { \bf E} ^ {\, r \,} (z, \omega) = \exp (i\kappa _ {\, 1 \,} z) [{\bf e} _
{\, l \,} C _ {\, T \,} (\omega)\nonumber\\
 + { \bf e} _ {\, l \,} ^ {*} C _ {\, T \,}
(-\omega)], \quad z\geq d,
\end{eqnarray}
and  the field vector to the left of the QW $ {\bf E} ^ {\, l \,} (z, \omega) $,
including the exciting pulse field Eq. (4) and the reflected wave field $ \Delta {\bf
E} ^ {\, l \,} (z, \omega) $ is equal
\begin{equation}
\label{26} { \bf E} ^ {\, l \,} (z, \omega) = {\bf E} _ {\, 0 \,} (z, \omega) +
\Delta {\bf E} ^ {\, l \,} (z, \omega),
\end{equation}
\begin{eqnarray}
\label{27} \Delta {\bf E} ^ {\, l \,} (z, \omega) = \exp- (i\kappa _ {\, 1 \,} z)
[{\bf e} _ {\, l \,} C _ {\, R \,} (\omega)\nonumber\\
 + { \bf e} _ {\, l \,} ^ {*} C
_ {\, R \,} (-\omega)], \quad z\geq d.
\end{eqnarray}

\section{Transition to time representation.}
In the time representation  the electrical field vector of the pulse transmitted
through the QW according to Eq. (17) is represented as $ (p=t-z \,\nu _ {\, 1 \,}/c) $
\begin{eqnarray}
\label{28} { \bf E} ^ {\, r \,} (z, t) = {\bf e} _ {\, l \,} E ^ {\, r \,} (z, t)
+c.c., \;\nonumber\\
 E^{\,r\,}(z,t)=(1/2\pi)\int_{-\infty}^{+\infty}d\omega\exp(-i\,\omega \,
 p)\nonumber\\
\times C _ {\, T \,} (\omega), \; z\geq d.
\end{eqnarray}
Similarly, the field vector  of the pulse, reflected from the QW, is equal
\begin{eqnarray}
\label{29} { \Delta \bf E} ^ {\, l \,} (z, t) = {\bf e} _ {\, l \,}\Delta E ^ {\, l
\,} (z, t) +c.c., \;\nonumber\\
 \Delta E ^ {\, l \,} (z, t) = (1/2\pi)
 \int_{-\infty}^{+\infty}d\omega\exp(-i\,\omega \, s)\nonumber\\
\times C _ {\, R \,} (\omega), \; z\leq 0,
\end{eqnarray}
where $s=t+z \,\nu _ {\, 1 \,}/c $, and the functions $C _ {\, T \,} (\omega) $ and $C
_ {\, R \,} (\omega) $ (after substitution in Eq. (18) $E _ {\, 0 \,} (\omega) $ from
Eq. (4) and $ {\cal N} (\omega) $ from Eq. (21)) look like
\begin{eqnarray}
\label{30}
 C_{\,T\,}(\omega)=\frac{4E_{\,0\,}\,\gamma_{\,l\,}\,\zeta\,
exp (-i\kappa _ {\, 1 \,} \, d)} {{\cal L} \ {, \cal D}} \;\nonumber\\
\times \frac {\omega - \omega_{\,0\,}-\gamma_{\,r\,}\varepsilon^{\,\prime\,\prime\,}/2
+i\gamma/2} {(\omega -\omega _ {\, l \,}) ^ 2 + (\gamma _ {\, l \,}/2) ^2},
\end{eqnarray}
\begin{eqnarray}
\label{31}
 C_{\,R\,}(\omega)=\frac{E_{\,0\,}\,\gamma_{\,l\,}}
{ {\cal L} \, {\cal D}}[ (\omega -\omega _ {\, l \,}) ^ 2 + (\gamma _ {\, l \,}/2) ^2]^{-2},
\nonumber\\
\times \{{\cal B} [\omega -\omega _ {\, 0 \,}-
 \gamma_{\,r\,}\varepsilon^{\,\prime\,\prime\,}/2+
 i(\gamma+\gamma_{\,r\,}\varepsilon^{\,\prime\,})/2]\nonumber\\-
i {\cal B}_{\,1\,}\gamma_{\,r\,}\varepsilon^{\,\prime\,}/2\},
\end{eqnarray}
\begin{equation}
\label{32} { \cal D} = \omega -\omega _ {\, 0 \,}-\gamma _ {\, r \,} {\cal F} _ {\, 1
\,}/2 + i (\gamma +\gamma _ {\, r \,} {\cal F} _ {\, 2 \,})/2,
\end{equation}
\begin{equation}
\label{33} { \cal L} = (1 +\zeta) ^2\exp (-i \,\kappa \, d) - (1-\zeta) ^2\exp (i
\,\kappa \, d),
\end{equation}
\begin{eqnarray}
\label{34} { \cal B} = -2 \, i \, (1-\zeta^2) \sin {\kappa \, d}, \nonumber\\
 { \cal
B}_{\,1\,}=2[1+\zeta^2-(1-\zeta^2)\exp(i\,\kappa \, d)],
\end{eqnarray}
\begin{eqnarray}
\label{35} { \cal F}_{\,1\,}=\varepsilon^{\,\prime\,\prime\,}-\frac {
2\,\varepsilon^{\,\prime\,}(1-\zeta^2)\sin{\kappa \, d}} { 1 +\zeta^2 + (1-\zeta^2)
\cos {\kappa \, d}} \;, \nonumber\\
 {\cal F} _ {\, 2 \,} =\frac {
2\,\zeta\,\varepsilon^{\,\prime\,}}{1+\zeta^2+(1-\zeta^2)\cos{\kappa \, d}} \;.
\end{eqnarray}
In integrals Eqs. (28), (29) poles of integrand functions are $ \omega =\omega _ {\,
l \,}\pm i\gamma _ {\, l \,}/2 $, and there is also the pole in the bottom
half-plane  $ \omega $, determined by the equation $ {\cal D} =0 $. Strictly
speaking, functions $ {\cal F} _ {\, 1 \,} $ and $ {\cal F} _ {\, 2 \,} $ included in
$ {\cal D} $ (Eq. (32)) depend from $ \omega $, since
 the module of a wave vector $ \kappa =\nu \, \omega/c $ depends from $ \omega $.
  However, by virtue of assumptions,
made at obtaining  Eq. (12)  $ \omega $ should not strongly differ from frequency $
\omega _ {\, 0 \,} $ and at the solution of the equation $ {\cal D} =0 $ it is enough
to be limited by the first iteration. It results in following poles in the bottom
half-plane
\begin{equation}
\label{36}
 \omega=\omega_{\,0\,}-\gamma_{\,r\,}{\cal F} _ {\, 1 \,} (\omega _ {\, 0 \,})-
i (\gamma + \gamma _ {\, r \,} {\cal F} _ {\, 2 \,} (\omega _ {\,
0 \,}))/2.
\end{equation}
At use of the approached value of the pole Eq. (36) we shall receive, that
\begin{equation}
\label{37} \kappa =\kappa _ {\, 0 \,} =\nu \, \omega _ {\, 0 \,}/c, \quad
 \kappa_{\,1\,}=\kappa_{\,1\,0\,}=\nu_{\,1\,} \omega _ {\, 0 \,}/c.
\end{equation}
On the other hand, the poles $ \omega =\omega _ {\, l \,}\pm i\gamma _ {\, l \,}/2 $
lead to $ \kappa =\kappa _ {\, l \,} =\nu \, \omega _ {\, l \,}/c $, $ \; $
 $\kappa_{\,1\,}=\kappa_{\,1\,l\,}=\nu_{\,1\,} \omega _ {\, l \,}/c $.
As the theory is true at performance of an inequality
 $(\omega_{\,l\,}-\omega_{\,0\,})/\omega_{\,0\,}\ll 1 $, further
we consider, that $\kappa_{\,l\,}=\kappa_{\,0\,}=\kappa,\,\kappa_{\,1\,l\,}= \kappa _
{\, 1 \, 0 \,} =\kappa _ {\, 1 \,} $.

After integration on $ \omega $ scalar functions $E ^ {\, r \,}
(z, t) $ and $ \Delta E ^ {\, l \,} (z, t) $ accept a kind
\begin{eqnarray}
\label{38} E ^ {\, r \,} (z, t) = (4\zeta E _ {\, 0 \,} / {\cal L}) \exp (-i (\omega
_ {\, l \,} p + \kappa _ {\, 1 \,} d))\nonumber\\
\times \{[1-\Theta(p)]\exp(\gamma_{\,l\,}p/2)W_{\,T\,}(\gamma_{\,l\,})+ \Theta (p)
\epsilon _ {\, T \,} \};
\end{eqnarray}
\begin{eqnarray}
\label{39} \Delta E ^ {\, l \,} (z, t) = (E _ {\, 0 \,} / {\cal L}) \exp (-i\omega _
{\, l \,} s)\nonumber\\
\times \{[1-\Theta(s)]\exp(\gamma_{\,l\,}s/2)W_{\,R\,}(\gamma_{\,l\,})+ \Theta (s)
\epsilon _ {\, R \,} \},
\end{eqnarray}
where the functions $ \epsilon_{T} $ and $ \epsilon_{R} $ are represented by the
uniform formula
\begin{eqnarray}
\label{40}
 \epsilon_{T(R)}=e^{-\gamma_{l}p(s)/2}
W_{T(R)}(-\gamma_l)\nonumber\\- e^{i(\Delta\omega-\gamma_{r}{\cal F}_{1}/2)p(s)}
W_{T(R)
}^\prime\nonumber\\
\times e^{-(\gamma +\gamma_{r}{\cal F}_{2})p(s)/2}.
\end{eqnarray}
In Eqs. (38) - (40) designations are entered
\begin{equation}
\label{41}
 \Delta\,\omega\,=\,\omega_{\,l\,}-\omega_{\,0\,}
\end{equation}
\begin{equation}
\label{42}
 W_{\,T\,}(\gamma_{\,l\,})=[\Delta\,\omega-\gamma_{\,r\,}\,
 \varepsilon^{\,\prime\,\prime\,}/2+i\,(\gamma+\gamma_{\,l\,})/2]/
\, \Omega (\gamma _ {\, l \,})
\end{equation}
\begin{eqnarray}
\label{43} W _ {\, R \,} (\gamma _ {\, l \,}) = \{{\cal B} [\Delta \,\omega-\gamma _
{\, r \,} \,
 \varepsilon^{\,\prime\,\prime\,}/2+i(\gamma + \gamma _ {\, l \,}\nonumber\\ +
\gamma _ {\, r \,} \,\varepsilon ^ {\,\prime \,}) /2] - i\gamma _ {\, r \,}
\,\varepsilon ^ {\,\prime \,} {\cal B} _ {\, 1 \,}/2 \}/\Omega (\gamma _ {\, l \,})
\end{eqnarray}
\begin{eqnarray}
\label{44} W _ {\, T \,} ^ {\,\prime \,} = -i \, (\gamma _ {\, r \,}/2) [ {\cal
F}_{\,2\,}-i(\varepsilon^{\,\prime\,\prime\,}-{\cal F} _ {\, 1 \,})]\nonumber\\
\times \left (\frac {1} {\Omega (-\gamma _ {\, l \,})}- \frac {1} {\Omega (\gamma _
{\, l \,})} \right)
\end{eqnarray}
\begin{eqnarray}
\label{45} W _ {\, R \,} ^ {\,\prime \,} = -i \, (\gamma _ {\, r \,}/2)\nonumber\\
\times\{{\cal B} [\varepsilon ^ {\,\prime \,} { -\cal
F}_{\,2\,}+i(\varepsilon^{\,\prime\,\prime\,}-{\cal F} _ {\, 1 \,})] + \varepsilon ^
{\,\prime \,} {\cal B} _ {\, 1 \,} \}\nonumber\\
\times \left (\frac {1} {\Omega (-\gamma _ {\, l \,})}- \frac {1} {\Omega (\gamma _
{\, l \,})} \right)
\end{eqnarray}
\begin{eqnarray}
\label{46}
 \Omega(\gamma_{\,l\,})=\Delta\,\omega-\gamma_{\,r\,}\,{\cal F} _ {\, 1 \,}/2\nonumber\\ +
 i(\gamma+\gamma_{\,l\,}+\gamma_{\,r\,}{\cal F} _ {\, 2 \,})/2.
\end{eqnarray}
Let us notice, that account of dependence $ \kappa $ from $ \omega $ results into
replacement in the expression for $E ^ {\, r \,} (z, t) $ (38) variable $p $ on $p ^
{\,\prime \,} = p+t _ {\, 1 \,} $, where $t _ {\, 1 \,} =\nu _ {\, 1 \,} \, d/c $
corresponds to time, during which light passes a QW. Thus, account of dependencies $
\kappa $ from $ \omega $ will have an effect only for $p\leq t _ {\, 1 \,} $. If
$d=500A, \, \nu _ {\, 1 \,} = 3 $, \, $t=5.10 ^ {-16} c. \,\cong t _ {\, 0 \,} $.
Since $t _ {\, 1 \,} \cong t _ {\, 0 \,} $, account of dependence $ \kappa $ from $
\omega $ at the calculation of Eqs. (28), (29) is excess of accuracy, as results in
amendments of the same order, which were not taken into account in Eq. (12). The
obtained expressions for $E ^ {\, r \,} (z, t) $ and $E ^ {\, l \,} (z, t) $ are very
large and their analytical research is complicated. Therefore two limiting cases
represent some interest, when these expressions  become essentially simpler. If the
matter is homogeneous, i.e. $ \nu_{\, 1 \,} =\nu $, then
$$\kappa _ {\, 1 \,} =\kappa, \: {\cal L} =4\exp (-i \,\kappa \, d), \: {\cal B} =0,
\:$$ $$ { \cal B} _ {\, 1 \,} = 4, \: {\cal
F}_{\,1\,}=\varepsilon^{\,\prime\,\prime\,}, \: {\cal F} _ {\, 2 \,} =\varepsilon ^
{\,\prime \,} $$ and Eqs. (38) and (39) pass in
\begin{eqnarray}
\label{47}  E ^ {\, r \,} (z, t) =E _ {\, 0 \,} (z, t)
+ \Delta E ^ {\, r \,} (z, t) =E _ {\, 0 \,} (z, t)\nonumber\\
 - E_{\,0\,}(i\,\gamma_{\,r\,}\,\varepsilon^{\,\prime\,}/2) \exp (-i
\,\omega _ {\, l \,} \, p) \,\nonumber\\
\times \{[1-\Theta (p)] \, \exp (\gamma _ {\, l \,} \, p/2) /\Omega (\gamma _ {\, l
\,}) + \Theta (p) \epsilon \},
\end{eqnarray}
\begin{eqnarray}
\label{48}
 \Delta E^{\,l\,}(z,t)=-E_{\,0\,}(i\gamma_{\,r\,}\,\varepsilon^{\,\prime\,}/2)
\exp (-i (\omega _ {\, l \,} \, s-\kappa d)) \nonumber\\
\times \{[1-\Theta (s)] \, \exp (\gamma _ {\, l \,} \, s/2) /\Omega (\gamma _ {\, l
\,}) + \Theta (s) \epsilon \},
\end{eqnarray}
where function $ \Omega (\gamma_{\, l \,}) $, determined in Eq. (40), turns in
\begin{equation}
\label{49} \Omega (\gamma _ {\, l \,}) = \Delta \,\omega-\gamma _ {\, r \,} \,
 \varepsilon^{\,\prime\,\prime\,}/2+i(\gamma + \gamma _ {\, l \,} +
\gamma _ {\, r \,} \,\varepsilon ^ {\,\prime \,}) /2
\end{equation}
and the function (40) accepts a kind
\begin{eqnarray}
\label{50} \epsilon = \exp (-\gamma _ {\, l \,} \, t/2) /\Omega (-\gamma _ {\, l
\,})\nonumber\\-
\exp[i(\Delta\,\omega-\gamma_{\,r\,}\,\varepsilon^{\,\prime\,\prime\,}/2)\,t]
 \exp[-(\gamma+\gamma_{\,r\,}\,\varepsilon^{\,\prime\,})\,t/2]\nonumber\\
\times\{\Omega (-\gamma _ {\, l \,} \}) ^ {-1} - \Omega (\gamma _ {\, l \,}) ^ {-1}
\}.
\end{eqnarray}
For $E^{\, r \,} $ parameter $t=p $, for $ \Delta \, E^{\, l \,} \; t=s $. Function $
\Delta E^ {\, r \,} (z, t) $ determines distortion  of an exciting pulse, transmitted
the QW.

It is seen from Eqs. (47) and (48) that the account of a  spatial dispersion in a case
of homogeneous matter results in a frequency  shift of $ \omega _ {\, 0 \,} $ on
magnitude $\gamma_{\,r\,}\,\varepsilon^{\,\prime\,\prime\,}/2$ and to replacement $
\gamma _ {\, r \,} $ by $ \tilde \gamma _ {\, r \,} =
 \gamma_{\,r\,}\,\varepsilon^{\,\prime\,}$.
The value $ \tilde \gamma _ {\, r \,} $ coincides with radiative broadenings of the
EHP in a strong magnetic field at $ \bf K _ {\,\perp \,} \, = 0 $ in case of any
value $ \kappa \, d $ calculated in \cite{3,7}. If the spatial dispersion is not
taken into account, i. e. $ \kappa d=0 $, then, according to Eq. (24), $ \varepsilon
^ {\,\prime \,}\to 1, \; \varepsilon ^ {\,\prime \,\prime \,}\to 0 $ and Eqs. (47)
and (48) pass into expressions obtained in \cite{5} for homogeneous matter in absence
of the spatial dispersion. Eqs. (47) and (48) coincide with similar expressions
obtained in \cite{5} (Eq. (15), if there under frequency transition $ \omega _ {\, 0
\,} $ to understand $ \omega _ {\, 0 \,} +
 \gamma_{\,r\,}\,\varepsilon^{\,\prime\,\prime\,}/2$, and under
$ \gamma _ {\, r \,} $ the value $ \tilde \gamma _ {\, r \,} $).

  A limiting case of a
weak spatial dispersion, when $ \kappa \, d \to 0 $, but the matter is inhomogeneous,
i. e. $ \nu _ {\, 1 \,}\neq \nu $ represents also some interest. It can take place for
comparatively narrow QWs. Believing in Eqs. (38) and (39) $ \kappa \, d=0 $, we shall
obtain, that $ {\cal L} =4 \,\zeta, \; {\cal B} =0, \: { \cal B} _ {\, 1 \,} = 4
\,\zeta^2, \; {\cal F} _ {\, 1 \,} = 0, \: {\cal F} _ {\, 2 \,} =\zeta $ and
\begin{eqnarray}
\label{51}
 \Delta\,E^{\,r\,}(z,t)=(-iE_{\,0\,}\,\gamma_{\,r\,}\,\zeta/2)
\exp (-i\omega _ {\, l \,} \, p)\nonumber\\
\times \{[1-\Theta (p)]
 \exp(\gamma_{\,l\,}\,p/2)/\Omega(\gamma_{\,l\,})+
\epsilon ^ {\,\prime \,} (p) \Theta (p) \}
\end{eqnarray}
\begin{eqnarray}
\label{52}
 \epsilon^{\,\prime\,}(p)=\exp(-\gamma_{\,l\,}\,p/2)/\Omega(-\gamma_{\,l\,})\nonumber\\-
\exp [i\Delta\,\omega\,p-(\gamma+\gamma_{\,r\,}\,\zeta)p/2]\nonumber\\
\times ( \Omega(-\gamma_{\,l\,})^{-1}-\Omega(\gamma_{\,l\,})^{-1}).
\end{eqnarray}
In this case
\begin{equation}
\label{53} \Omega (\gamma _ {\, l \,}) =\Delta \,\omega +
 i(\gamma+\gamma_{\,l\,}+\gamma_{\,r\,}\zeta/2),
\end{equation}
and $ \Delta E ^ {\, l \,} (z, t) $ differs from Eq. (51) by the replacement $p $ by
$s $. One can see, that the matter  heterogeneity  without taking into account the
spatial dispersion results only into replacement $ \gamma _ {\, r \,} $ by $ \gamma _
{\, r \,} \,\zeta $, i. e. to the replacement in Eq. (8) for $ \gamma _ {\, r \,}
\;\nu $ on $ \nu _ {\, 1 \,} $. Eqs. (51) and (52) coincide with obtained in
\cite{5}, if $ \gamma _ {\, r \,} $ there is  replaced by $ \zeta \,\gamma _ {\, r
\,} $. Since in real systems $ \zeta \cong 1 $, the changes, which are brought in
only by heterogeneity of matter, are insignificant. The limit transition $ \gamma _
{\, l \,}\to 0 $ means a transition to a monochromatic exciting wave. In this
limiting case Eqs. (38) and (39) pass in expressions obtained  in \cite{11}.

\section{ Reflection and transmission of an exciting pulse.}

The energy flux $ {\bf S} (p) $, appropriate to the electrical field of the exciting
pulse, is equal
\begin{equation}
\label{54} { \bf S} (p) = ({\bf e} _ {\, z \,}/4 \,\pi) (c/\nu _ {\, 1 \,}) ({\bf E}
_ {\, 0 \,} \ (, z, t)) ^2 = { \bf e} _ {\, z \,} S _ {\, 0 \,} P (p),
\end{equation}
where $S _ {\, 0 \,} = cE _ {\, 0 \,} ^ 2 / (2\pi \nu _ {\, 1 \,}) $, $ {\bf e} _ {\,
z \,} $ is the unit vector in a direction $z $. The dimensionless function $P (p) $
determines the spatial and time dependence of the energy flux of the exciting pulse
\begin{eqnarray}
\label{55} P (p) = ({\bf E} _ {\, 0 \,} (z, t)) ^2/S _ {\, 0 \,} =\Theta (p) e ^
{-\gamma _ {\, l \,} \, p}\nonumber\\ + [ 1-\Theta (p)] e ^ {\gamma _ {\, l \,} \, p}.
\end{eqnarray}
The transmitted flux to the right of the QW, by analogy with Eq. (54), looks like
\begin{eqnarray}
\label{56} { \bf S} ^ {\, r \,} \ (, z, t) = ({\bf e} _ {\, z \,}/4 \,\pi) (c/\nu _
{\, 1 \,}) ( {\bf E} ^ {\, r \,} \, (z, t)) ^2\nonumber\\ = {\bf e} _ {\, z \,} S _
{\, 0 \,} {\cal T} (p).
\end{eqnarray}
For the reflected flux (to the left of a QW) we obtain
\begin{eqnarray}
\label{57} { \bf S} ^ {\, l \,} =- ({\bf e} _ {\, z \,}/4 \,\pi) (c/\nu _ {\, 1 \,})
( {\Delta \bf E} ^ {\, l \,} \, (z, t)) ^2\nonumber\\ =- {\bf e} _ {\, z \,} S _ {\, 0
\,} {\cal R} (s).
\end{eqnarray}
The dimensionless functions $ {\cal T} (p) $ and $ {\cal R} (s) $ determine the shares
of the transmitted and reflected energy of the exciting pulse.

Let us define, by analogy with \cite{5}, the  absorbed energy flux $ {\bf S} ^ {\, a
\,} $ as a difference of the flux   $ {\bf S} \, + \, {\bf S} ^ {\, l \,} $ at $z=0 $
entered in the QW from the left, and flux ,  $ {\bf S} ^ {\, r \,} $ at $z=d $
leaving the QW, on the right at the same moment of time $t $:
\begin{equation}
\label{58} { \bf S} ^ {\, a \,} \ (, t) = {\bf S} (t) + {\bf S} ^ {\, l \,} \, (t) -
{\bf S} ^ {\, r \,} \, (t).
\end{equation}
Using the definitions Eqs. (54) - (58) let us present $ {\bf S} ^ {\, a \,} \, (t) $
as
\begin{equation}
\label{59} { \bf S} ^ {\, a \,} \ (, t) = {\bf e} _ {\, z \,} \, S _ {\, 0 \,} \ [, P
(t) - {\cal R} (t) - { \cal T} (t)].
\end{equation}
Defining a share of the absorbed energy $ {\cal A} (t) $ by equality $ {\bf S} ^ {\,
a \,} \, (t) = {\bf e} _ {\, z \,} \, S _ {\, 0 \,} \, {\cal A} (t) $ we shall obtain,
that
\begin{equation}
\label{60} { \cal A} (t) =P (t) - {\cal R} (t) - {\cal T} (t).
\end{equation}
The Eq. (60) can be generalized if to remove planes, in which are observed flows on
distance $z =-z _ {\, 0 \,} $ (to the left of well) and on $z _ {\, 0 \,} $ to the
right of a QW $z _ {\, 0 \,} > 0 $. Then instead of Eq. (60) we shall obtain
\begin{equation}
\label{61} { \cal A} (x) =P (x) - {\cal R} (x) - {\cal T} (x),
\end{equation}
where $x=p=s=t-\nu _ {\, 1 \,} \, | z _ {\, 0 \,} \, |/c $. Expressions for the values
$ {\cal T} $, $ {\cal R} $ and $ {\cal A} $, which are determined by scalars $E ^ {\,
r \,} (z, t) $ and $ \Delta E ^ {\, l \,} (z, t) $ on general Eqs. (38) and (39) are
not shown here in view of their extreme size. The values $P (t) $, $ {\cal T} (t) $
and $ {\cal R} (t) $ are always positive, $ {\cal A} (t) $ can be of any sign. The
negative absorption at some moment of time $t $ means, that the electronic system of
a QW gives back energy saved in previous moments of time.

\section{ Time dependence of reflection, transmission and absorption in
the resonance $ \omega _ {\, l \,} =\omega _ {\, 0 \,}. $}

Let's consider at first the limiting case $ \gamma\gg\gamma _ {\, r \,} $. Then fields
$ {\bf E} ^ {\, r \,} (z, t) $ and $ \Delta {\bf E} ^ {\, l \,} (z, t) $ from Eqs.
(38) and (39) can be represented as a decomposition to a row
\begin{equation}
\label{62} { \bf E} ^ {\, r \,} (z, t) = {\bf E} _ {\, 0 \,} ^ {\, r \,} (z, t) + (
\gamma _ {\, r \,}/\gamma) {\bf E} _ {\, 1 \,} ^ {\, r \,} (z, t) +...,
\end{equation}
\begin{equation}
\label{63} \Delta {\bf E} ^ {\, l \,} (z, t) = \Delta {\bf E} _ {\, 0 \,} ^ {\, l \,}
(z, t) + ( \gamma _ {\, r \,}/\gamma) \Delta {\bf E} _ {\, 1 \,} ^ {\, l \,} (z, t)
+...,
\end{equation}
where
\begin{eqnarray}
\label{64}  {\bf E} _ {\, 0 \,} ^ {\, r \,} (z, t) = {\bf e} _ {\, l \,} (4 \,\zeta
\, E _ {\, 0 \,} / {\cal L}) \exp [-
i(\omega_{\,l\,}p+\kappa_{\,1\,}d)]\nonumber\\
\times
 \{[1-\Theta (p)] \exp (\gamma _ {\, l \,} \, p/2) + \Theta (p) \exp
(-\gamma _ {\, l \,} \, p/2) \}\nonumber\\ +c.c.,
\end{eqnarray}
\begin{eqnarray}
\label{65} \Delta {\bf E} _ {\, 0 \,} ^ {\, l \,} (z, t) = - {\bf e} _ {\, l \,} (
{\cal B} E _ {\, 0 \,} / {\cal L}) \exp (-i \,\omega _ {\, l \,} \, s)\nonumber\\
\times
 \{[1-\Theta (s)] \exp (\gamma _ {\, l \,} \, s/2) + \Theta (s) \exp
(-\gamma _ {\, l \,} \, s/2) \}\nonumber\\ +c.c.
\end{eqnarray}
correspond to fields of transmitted and reflected pulses at $ \gamma_{\, r \,} = 0 $,
i. e. when the absorption in a QW is absent.

In limiting cases $ \kappa \, d\neq 0, \; \zeta =1 $ or $ \kappa \, d=0, \; \zeta
\neq 1 \quad \Delta E _ {\, 0 \,} ^ {\, l \,} (z, t) =0 $, since, according to Eq.
(34), $ {\cal B} =0 $. In the first case it is connected with that that the matter
becomes homogeneous, in the second case - that a substance in QW becomes very little
and the transmitting wave does not react to it. In these limiting cases, as it
follows from Eq. (57), $ {\cal R} (t) \sim (\gamma _ {\, r \,}/\gamma) ^2 $, i. e. is
a small value. At transition to a general case $ \kappa \, d\neq 0, \; \zeta \neq 1
\; {\cal R} (t) $ looks like
\begin{eqnarray}
\label{66} { \cal R} (t) =S _ {\, 0 \,} ^ {-1} [(\Delta {\bf E} _ {\, 0 \,} ^ {\, l
\,} (s)) ^2\nonumber\\ + 2 (\gamma _ {\, r \,}/\gamma) (\Delta {\bf E} _ {\, 0 \,} ^
{\, l \,} (s) \, \Delta {\bf E} _ {\, 1 \,} ^ {\, l \,} (s))],
\end{eqnarray}
what will result in essential increase of reflection at the expense of first term in
Eq. (66). As to transmission $ {\cal T} (t) $, in limiting cases $ \kappa d=0 $ or $
\zeta =1 \; {\cal T} (t) =P (t) $. At transition to a general case transmission
changes poorly, as a multiplier $16\zeta / | {\cal L} | ^2 $  here does not too
strongly differ from unit.

 In Fig. 1 the time  dependencies of the dimensionless transmission
 $ {\cal T} $, absorption $ {\cal A} $ and reflection $ {\cal R} $ are represented
  for different
 values of parameters $ \kappa \, d $ and $ \zeta $. In Fig. 1  it is visible, that curves
 $ {\cal T} (t) $ practically coincide for $ \kappa \, d = 0, \;
 \zeta = 1 $ and $ \kappa \, d=1.5, \, \zeta = 1.1 $. The same takes
place for absorption $ {\cal A} (t) $. On the other hand, as it is visible in Fig. 1b,
the reflection, being a small value, essentially depends on parameter
 $ \zeta $ for $ \kappa \, d=1.5 $: at change $ \zeta $ from 1 up to 1.3 $ {\cal R} (t) $
 grows in 8 times.

 In a limiting case $ \gamma _ {\, r \,}\gg \gamma $ the induced fields are comparable
 on value with a field of an exciting pulse and consequently the form of the
 transmitted
 through a QW  a pulse varies very strongly. It is visible in Fig. 2,
where transmission is small, and the reflection $ {\cal R} $ dominates.
 In \cite{5} the concept of the special points on time curves
 $ {\cal T} $, $ {\cal A} $ and $ {\cal R} $ was entered. In particular, one of these points
 (the point of total reflection of the first type) was determined by a condition
 $ {\cal R} (t _ {\, 0 \,}) = P (t _ {\, 0 \,}), \; {\cal T} (t _ {\, 0 \,}) =
 {\cal A} (t _ {\, 0 \,}) = 0 $
 (see Fig. 2 ). If to take into account, that $ \zeta\neq 1, \; \kappa \, d\neq 0 $ (fig. 2b),
then in a point of total reflection  other condition takes place, namely
 $ {\cal T} (t _ {\, 0 \,}) + {\cal A} (t _ {\, 0 \,}) = 0, \quad
 {\cal R} (t _ {\, 0 \,}) = P (t _ {\, 0 \,}) $. It means,
 that $ {\cal A} (t _ {\, 0 \,}) < 0 $, i. e. the generation of radiation takes place,
which was saved by system at earlier moments of time.
 Arisen at transition to a general case the special point according to the classification
 of \cite{5}
 is the special point of total reflection of the second type. Let's note also, that
 in Fig. 2b transmission in some times is more, than in Fig. 2 , i. e. and in
this case the heterogeneity of matter and spatial dispersion strongly influence only
on small values, which in the given limiting case is $ {\cal T} (t) $.

 \section{ A deviation of carrying frequency from resonant.}

 In \cite{5} it was shown, that the carrying frequency deviation $ \Delta \, \omega $
  results into time oscillations of $ {\cal A} (t) $
and $ {\cal R} (t) $. However, oscillations were distinct only for small values. On
the other hand, taking into account  the matter heterogeneity
  and spatial dispersion results in occurrence additional
reflections from borders of a QW, which can exceed the oscillating component $ {\cal
R} (t) $. In Fig. 3 the examples of influence of the matter heterogeneity and spatial
dispersion of light wave on
 function $ {\cal R} (t) $ are given.
The  most significant changes take place in the  case $ \gamma _ {\, r \,}/\gamma _
{\, l \,}\ll 1 $, i. e. for the short exciting pulse. It is visible in Fig. 3a the
value $ {\cal R} (0) $
 is increased in comparison with a case $ \zeta =1, \:\kappa \, d=0 $ more
 than in 300 times, oscillations here are indiscernible in view of their small amplitude.
 For an intermediate case $ \gamma _ {\, r \,} =\gamma _ {\, l \,} $ (Fig. 3b)
 the changes are insignificant and oscillations  on a curve,
corresponding to $ \zeta =1.1, \:\kappa d=1.5 $ are well visible.
 In  Fig. 3á reflection $ {\cal R} (0) $ is increased in 22 times, oscillations
 are still distinct. As to absorption, oscillating curves
$ {\cal A} (t) $ poorly change at transition to an inhomogeneous matter. It
 is explained by the fact that  absorption is caused by quantum transitions in
  QW, which poorly depend on a refraction index.

 In summary in Fig. 4 the curves $ {\cal R} (t) $ for a case
 $ \gamma _ {\, r \,}\gg \gamma _ {\, l \,} $ are given (a long exciting pulse),
 when $ \Delta \, \omega \neq 0 $, but oscillations of reflection practically
 are imperceptible. It follows from Fig. 4 that the account only of the spatial dispersion,
 reduces reflection in comparison with a case $ \kappa \, d=0 $. It is explained
 by reduction of radiative broadening
 $\gamma_{\,r\,}\,\varepsilon^{\,\prime\,}$, since
 $ \varepsilon ^ {\,\prime \,} $ is a decreasing function of parameter $ \kappa \, d $.
 The transition to inhomogeneous matter results in increasing of reflection by that greater,
 than more the parameter $ \zeta $.

On the basis of obtained results it is possible to make a general conclusion, that
the account of the matter heterogeneity  and spatial dispersion of plane waves,
consisting the exciting pulse, influences strongly only reflection. Changes are most
significant in that case, when
 reflection, connected with interband transitions in a QW, is small
 and it
masks by stronger reflection from borders of a QW. It takes place in the limiting case
$ \gamma\gg \gamma _ {\, r \,} $ (it was demonstrated above for the exact resonance $
\Delta \, \omega=0 $) and at a deviation of a carrying frequency from resonant in the
other limiting case
 $ \gamma \ll \gamma _ {\, r \,} $. It is necessary to note also strong dependence
of reflection from the parameter $ \nu /\nu _ {\, 1 \,} $, which is increased for the
account
 reflections from borders of a QW. The change of transmission also takes place only
 if it is not small.

 In real semiconductor heterostructures the impure electrons of a barrier
 transit   in a QW, deforming near to borders its rectangular form.
 Therefore advanced above theory is true for pure substances and
 wide QWs, when the size of the deformed frontier areas is small in comparison
 to a QW width. Besides the theory is true for deep QWs,
 the position of first energy levels in which and wave functions, appropriate to
 them,
 differ weakly from the position of energy levels and wave functions in indefinitely
 deep QWs. Since in the theory one exited level is taken into account only,
 the neighbour energy levels in a QW should be located further,
 than broadening of a considered energy level, and broadening of an exciting
light pulse should be less than distances between neighbour  levels.
 These requirements impose restriction from above on a QW width. For example,
 for $d=500A $ and $m _ {\, c \,} = 0.06 \, m _ {\, 0 \,} $ the difference of two
  lowest size-quantized energy levels
  is $ \cong 10 ^ {-3} eV $.

\section{Acknowledgements}
 The work is
executed at financial support of RFBI (project ü 00-02-16904) and Program " Physics
of solid-state nanostructures (97-1099). S. '. Pavlov thanks University of Zacatecas
for financial support and hospitality. D. A. Contreras-Solorio thanks CONACyT
(27736-E) for financial support.

\begin{figure}
\caption{Fig. 1. Transmission $ {\cal T} $, absorption $ {\cal A} $ and reflection $
{\cal R} $ of a symmetric exciting pulse $P $ as function of dimensionless time $
\gamma _ {\, l \,} \, t $. A resonant case, $ \gamma _ {\, r \,}\ll \gamma $. a: a
curve 1 - $ \zeta =1, \, \kappa \, d=0 $ and $ \zeta =1.1, \, \kappa \, d=1.5 $, 2 - $
\zeta =1, \, \kappa \, d=0 $, 3 - $ \zeta =1.1, \, \kappa \, d=1.5 $, b: 1 - $ \zeta
=1, \, \kappa \, d=0 $, 2 - $ \zeta =1, \, \kappa \, d=1.5 $, 3 - $ \zeta =1.1, \,
\kappa \, d=1.5 $, 4 - $ \zeta =1.2, \, \kappa \, d=1.5 $, 5 - $ \zeta =1.3, \, \kappa
\, d=1.5 $.}
\end{figure}

\begin{figure}
\caption{ Fig. 2. Time dependence $P, {\cal R}, {\cal A} $ and $ {\cal T} $ in the
resonant case, $ \gamma _ {\, r \,}\gg \gamma $. a - $ \zeta =1, \, \kappa \, d=0 $,
b - $ \zeta =1.1, \, \kappa \, d=1.5 $, $ \gamma _ {\, l \,} t $ and $ \gamma _ {\, l
\,} t _ {\, r \,} $ - special points of total reflections 1 and 2,respectively.}
\end{figure}

\begin{figure}
\caption{ Fig. 3. Time dependence of reflection in case of a deviation carrying
frequency from resonant. a, b, c:
 Curve 1 - $ \zeta =1, \, \kappa \, d=0 $, \, 2 - $ \zeta =1.1, \, \kappa \, d=1.5 $.
}
\end{figure}

\begin{figure}
\caption{ Fig. 4. Reflection in the case of strong deviation of carrying frequency
from resonant. Curve 1 - $ \zeta =1, \, \kappa \, d=0 $, \, 2 - $ \zeta =1.1, \,
\kappa \, d=1.5 $, 3 - $ \zeta =1.2, \, \kappa \, d=1.5 $, 4 - $ \zeta =1.3, \, \kappa
\, d=1.5 $.}
\end{figure}
\end{document}